\documentclass[aps,pra,superscriptaddress]{revtex4}
\usepackage{amssymb}
\usepackage{graphicx}
\usepackage{dcolumn}
\usepackage{bm}

\begin{document}

\title{Interference effects between two initially independent
Bose-condensed gases}
\author{Hongwei Xiong}
\affiliation{State Key Laboratory of Magnetic Resonance and Atomic and Molecular Physics,
Wuhan Institute of Physics and Mathematics, Chinese Academy of Sciences,
Wuhan 430071, P. R. China}
\affiliation{Center for Cold Atom Physics, Chinese Academy of Sciences,\\
Wuhan 430071, P. R. China}
\affiliation{Graduate School of the Chinese Academy of Sciences, Wuhan 430071, P. R. China}
\author{Shujuan Liu}
\affiliation{State Key Laboratory of Magnetic Resonance and Atomic and Molecular Physics,
Wuhan Institute of Physics and Mathematics, Chinese Academy of Sciences,
Wuhan 430071, P. R. China}
\affiliation{Center for Cold Atom Physics, Chinese Academy of Sciences,\\
Wuhan 430071, P. R. China}
\affiliation{Graduate School of the Chinese Academy of Sciences, Wuhan 430071, P. R. China}
\date{\today }

\begin{abstract}
When two initially independent Bose-condensed gases are allowed to overlap,
we investigate the density expectation value of the whole system by using
the second quantization method. In the presence of interatomic interaction,
based on the exact expression of the density expectation value, it is found
that there is a nonzero interference term in the density expectation value
of the whole system. The evolution of the density expectation value is shown
for different coupling constants. The present work shows clearly that there
is an interaction-induced coherence process between two initially
independent condensates.

\end{abstract}

\maketitle

\subsection{\protect\bigskip Introduction}

After the experimental realization of Bose-Einstein condensates about ten
years ago \cite{Anderson,Davis,Bradley}, the intensive experimental and
theoretical studies have shown clearly that the Bose condensate has stable
coherence property \cite{rev1,rev2,rev3}. After two separated condensates
are allowed to overlap, the coherence property of the whole system is a very
interesting problem. Interference patterns between Bose condensates can give
us important information about the first-order coherence properties of the
whole system. Thus, in the celebrated experiment by Andrews \textit{et al}
\cite{Andrew}, the interference patterns between two separated condensates
are investigated. In this experiment, the interference patterns between two
coherently separated Bose condensates were observed clearly. In the same
experiment, there is also a quite striking experimental result that there
are high-contrast fringes even for two completely independent condensates at
an initial time.

In our recent theoretical paper \cite{Xiong}, based on the exact expression
of the density expectation value calculated from the many-body wavefunction
of the whole system, it is shown clearly that there is an
interaction-induced coherence process when two initially independent
condensates are allowed to overlap. Due to the interaction-induced coherence
process between two initially independent condensates, there would be
interference patterns in the density expectation value when there is an
overlapping between two initially independent condensates. In the present
work, we will investigate the density expectation value based on the second
quantization method. It is shown that the result calculated from the second
quantization method is the same as the result based on the many-body
wavefunction. The calculations based on the second quantization method will
give us further physical picture for the interaction-induced coherence
process. In particular, to show clearly the interaction-induced coherence
process, we calculate the density expectation value for different coupling
constants in this work. It is shown clearly that increasing the interatomic
interaction will enhance the coherence effect of the whole system.

\subsection{The exact expression of the density expectation value}

Two separated condensates can be created by a double-well trap
(See for example the experiments of Refs
\cite{Andrew,double-well}). When the tunneling effect between two
wells can be omitted, one can obtain two completely independent
condensates with laser cooling and the following evaporative
cooling in the presence of the double-well trap. Obviously, when
the tunneling effect between two wells is obvious, the two
condensates confined in the double-well trap are coherently
separated. The interference patterns of the whole system can be
investigated by switching off the double-well trap. As shown in
Fig.1(a), there is no overlapping between two initially
independent condensates. After the confined potential is switched
off, two condensates will expand freely and result in the
overlapping.

\begin{figure}[tbp]
\includegraphics[width=0.6\linewidth,angle=270]{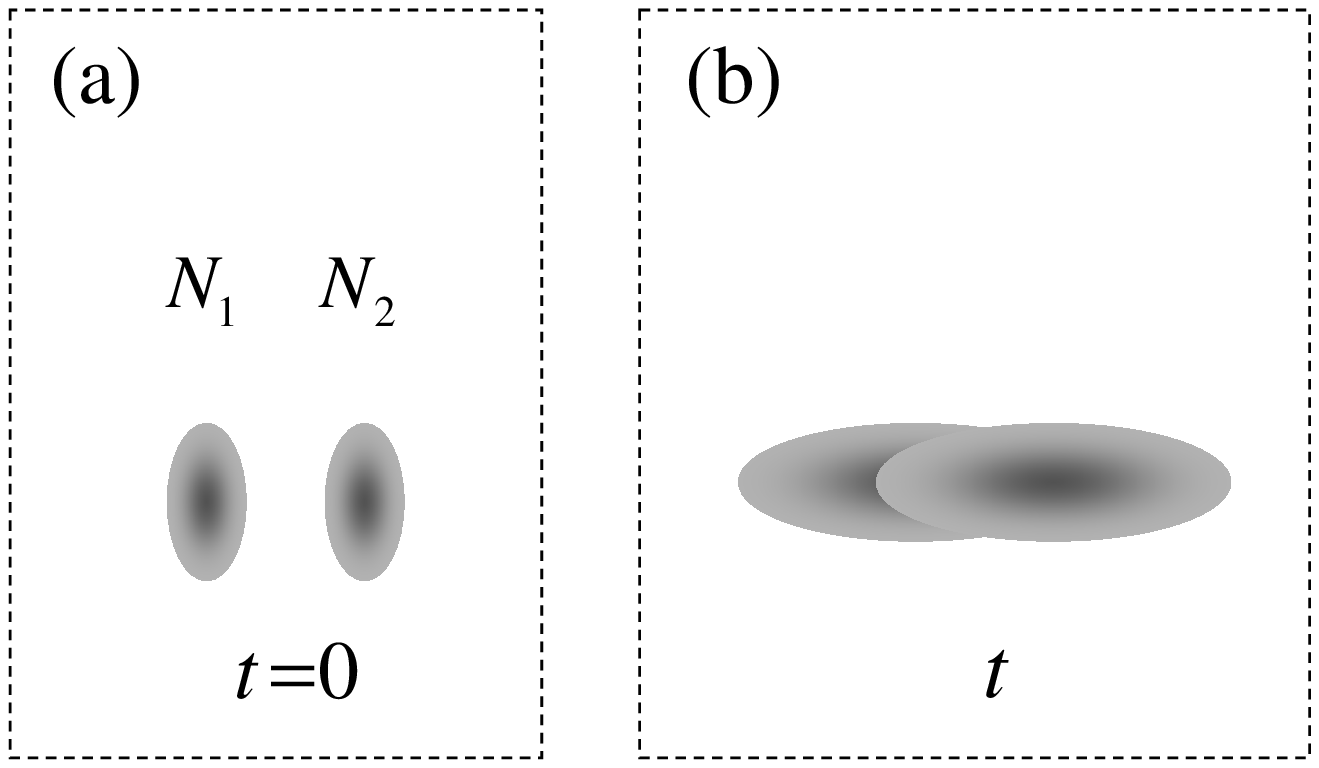}
\caption{Figure 1(a) shows two initially independent condensates,
while figure 1(b) shows the overlapping of two condensates due to
free expansion.}
\end{figure}

For two initially independent condensates comprising particle number $N_{1}$
and $N_{2}$, the corresponding state is (see also \cite{Pethick,Leggett}):%
\begin{equation}
\left\vert N_{1},N_{2}\right\rangle =\frac{C_{n}}{\sqrt{N_{1}!N_{2}!}}(%
\widehat{a}_{1}^{\dag })^{N_{1}}(\widehat{a}_{2}^{\dag })^{N_{2}}\left\vert
0\right\rangle ,  \label{initial-state}
\end{equation}%
where $C_{n}$ is a normalization constant to assure $\left\langle
N_{1},N_{2}|N_{1},N_{2}\right\rangle =1$. $\widehat{a}_{1}^{\dag }$ ($%
\widehat{a}_{2}^{\dag }$) is a creation operator which creates a particle
described by the single-particle state $\phi _{1}$ ($\phi _{2}$) in the left
(right) condensate.

One should note that for two initially coherently-separated condensates, the
state is%
\begin{equation}
\left\vert N\right\rangle =\frac{1}{\sqrt{N!}}(\widehat{b}^{\dagger
})^{N}\left\vert 0\right\rangle ,  \label{coherently-separated}
\end{equation}%
where $\widehat{b}^{\dagger }$ is a creation operator which creates a
particle with the single-particle state $\left( \sqrt{N_{1}}\phi _{1}+\sqrt{%
N_{2}}\phi _{2}\right) /\sqrt{N_{1}+N_{2}}$. For two coherently-separated
condensates, the density expectation value is
\begin{eqnarray}
n_{c}\left( \mathbf{r},t\right) &=&\left\langle N,t\right\vert \widehat{\Psi
}^{\dag }\left( \mathbf{r},t\right) \widehat{\Psi }\left( \mathbf{r}%
,t\right) \left\vert N,t\right\rangle  \nonumber \\
&=&N\left[ a_{c}\left\vert \phi _{1}\left( \mathbf{r},t\right) \right\vert
^{2}+2b_{c}\times \mathrm{Re}\left( \phi _{1}^{\ast }\left( \mathbf{r}%
,t\right) \phi _{2}\left( \mathbf{r},t\right) \right) +c_{c}\left\vert \phi
_{2}\left( \mathbf{r},t\right) \right\vert ^{2}\right] .  \label{coh-density}
\end{eqnarray}%
where $a_{c}=N_{1}/N$, $b_{c}=\sqrt{N_{1}N_{2}}/N$ and $c_{c}=N_{2}/N$ with $%
N=N_{1}+N_{2}$. The second term in the above equation accounts for the
interference effect when there is an overlapping between two condensates
upon expansion. For two coherently-separated condensates, the interference
patterns were investigated in several interesting theoretical works \cite%
{Rohrl,Liu}.

For two initially independent condensates, however, for the state given by
Eq. (\ref{initial-state}), it seems that there is no interference term in
the density expectation value with the following simple calculation (See
also \cite{Pethick,Leggett}):
\begin{eqnarray}
n_{d}\left( \mathbf{r},t\right) &=&\left\langle N_{1},N_{2},t\right\vert
\widehat{\Psi }^{\dag }\left( \mathbf{r},t\right) \widehat{\Psi }\left(
\mathbf{r},t\right) \left\vert N_{1},N_{2},t\right\rangle
\label{nd-field(1)} \\
&=&\left\langle N_{1},N_{2},t\right\vert \left( \widehat{a}_{1}^{\dag }%
\widehat{a}_{1}\left\vert \phi _{1}\left( \mathbf{r},t\right) \right\vert
^{2}+\widehat{a}_{2}^{\dag }\widehat{a}_{2}\left\vert \phi _{2}\left(
\mathbf{r},t\right) \right\vert ^{2}\right) \left\vert
N_{1},N_{2},t\right\rangle  \nonumber \\
&&+2\times \mathrm{Re}\left( \left\langle N_{1},N_{2},t\right\vert \widehat{a%
}_{1}^{\dag }\widehat{a}_{2}\left\vert N_{1},N_{2},t\right\rangle \phi
_{1}^{\ast }\left( \mathbf{r},t\right) \phi _{2}\left( \mathbf{r},t\right)
\right)  \label{nd-center} \\
&=&N_{1}\left\vert \phi _{1}\left( \mathbf{r},t\right) \right\vert
^{2}+N_{2}\left\vert \phi _{2}\left( \mathbf{r},t\right) \right\vert ^{2}.
\label{nd-field(2)}
\end{eqnarray}%
In the above equation, the field operator $\widehat{\Psi }\left( \mathbf{r}%
,t\right) $ is expanded as $\widehat{\Psi }\left( \mathbf{r},t\right) =%
\widehat{a}_{1}\phi _{1}+\widehat{a}_{2}\phi _{2}+\cdots $ with $\widehat{a}%
_{1}$ and $\widehat{a}_{2}$ being the annihilation operators. One should
note that, to get (\ref{nd-field(2)}) from (\ref{nd-field(1)}), there is an
implicit assumption that $\widehat{a}_{1}$ and $\widehat{a}_{2}^{\dag }$ are
commutative. This holds when $\int \phi _{1}\left( \mathbf{r},t\right) \phi
_{2}^{\ast }\left( \mathbf{r},t\right) dV=0$. When $\left[ \widehat{a}_{1},%
\widehat{a}_{2}^{\dag }\right] =0$, it is easy to understand that the
interference term (the last term in Eq. (\ref{nd-center})) is zero in $%
n_{d}\left( \mathbf{r},t\right) $.

In the present work, however, we will show that $\phi _{1}$ and $\phi _{2}$
will become non-orthogonal in the presence of interatomic interaction. Thus,
we will investigate the general case for $\int \phi _{1}\phi _{2}^{\ast
}dV=\zeta $. The operators $\widehat{a}_{1}$ and $\widehat{a}_{2}$ can be
written as%
\begin{equation}
\widehat{a}_{1}=\int \widehat{\Psi }\phi _{1}^{\ast }dV,  \label{a1}
\end{equation}%
and%
\begin{equation}
\widehat{a}_{2}=\int \widehat{\Psi }\phi _{2}^{\ast }dV.  \label{a2}
\end{equation}%
Here $\widehat{\Psi }\left( \mathbf{x},t\right) $ is the field operator. By
using the commutation relations of the field operators $[\widehat{\Psi }%
\left( \mathbf{x},t\right) ,\widehat{\Psi }\left( \mathbf{y},t\right) ]=0$
and $[\widehat{\Psi }\left( \mathbf{x},t\right) ,\widehat{\Psi }^{\dagger
}\left( \mathbf{y},t\right) ]=\delta \left( \mathbf{x}-\mathbf{y}\right) $,
it is easy to get the commutation relation%
\begin{equation}
\lbrack \widehat{a}_{1},\widehat{a}_{2}^{\dagger }]=\zeta ^{\ast }.
\label{commre}
\end{equation}%
We see that $\widehat{a}_{1}$ and $\widehat{a}_{2}^{\dagger }$ are not
commutative any more for $\int \phi _{1}\phi _{2}^{\ast }dV$ being a nonzero
value. In this case, it is obvious that one can not get the result (\ref%
{nd-field(2)}) from (\ref{nd-field(1)}). This means that one should be very
careful to get the correct density expectation value for two initially
independent condensates.

It is well-known that the field operator should be expanded in terms of a
complete and orthogonal basis set. Generally speaking, the field operator $%
\widehat{\Psi }\left( \mathbf{r},t\right) $ can be expanded as:
\begin{equation}
\widehat{\Psi }\left( \mathbf{r},t\right) =\widehat{a}_{1}\phi _{1}\left(
\mathbf{r},t\right) +\widehat{k}\phi _{2}^{\prime }\left( \mathbf{r}%
,t\right) +\cdots ,  \label{new-expansion}
\end{equation}%
where $\phi _{1}$ and $\phi _{2}^{\prime }$ are orthogonal normalization
wavefunctions. Assuming that $\phi _{2}^{\prime }=\beta \left( \phi
_{2}+\alpha \phi _{1}\right) $, based on the conditions $\int \phi
_{1}^{\ast }\phi _{2}^{\prime }dV=0$ and $\int \left\vert \phi _{2}^{\prime
}\right\vert ^{2}dV=1$, we have $\left\vert \beta \right\vert =\left(
1-\left\vert \zeta \right\vert ^{2}\right) ^{-1/2}$ and $\alpha =-$ $\zeta
^{\ast }$. Based on%
\begin{equation}
\widehat{k}=\int \widehat{\Psi }\left( \phi _{2}^{\prime }\right) ^{\ast }dV,
\label{k-operator}
\end{equation}%
we have%
\begin{equation}
\widehat{a}_{2}=\widehat{k}/\beta ^{\ast }+\zeta \widehat{a}_{1}.
\label{a2-k}
\end{equation}%
It is easy to get the following commutation relations:
\begin{eqnarray}
\lbrack \widehat{k},\widehat{k}] &=&[\widehat{k}^{\dagger },\widehat{k}%
^{\dagger }]=0,[\widehat{k},\widehat{k}^{\dagger }]=1,  \nonumber \\
\left[ \widehat{a}_{1},\widehat{a}_{1}\right] &=&[\widehat{a}_{1}^{\dagger },%
\widehat{a}_{1}^{\dagger }]=0,[\widehat{a}_{1},\widehat{a}_{1}^{\dagger }]=1,
\nonumber \\
\lbrack \widehat{k},\widehat{a}_{1}] &=&[\widehat{k},\widehat{a}%
_{1}^{\dagger }]=0.  \label{kcom}
\end{eqnarray}

Because $\widehat{k}$ and $\widehat{a}_{1}^{\dagger }$ are commutative, it
is convenient to calculate the density expectation value $n_{d}\left(
\mathbf{r},t\right) $ by using the operators $\widehat{k}$ and $\widehat{a}%
_{1}^{\dagger }$. The exact expression of the density expectation value is
\begin{eqnarray}
n_{d}\left( \mathbf{r},t\right) &=&\left\langle N_{1},N_{2},t\right\vert
\widehat{\Psi }^{\dag }\left( \mathbf{r},t\right) \widehat{\Psi }\left(
\mathbf{r},t\right) \left\vert N_{1},N_{2},t\right\rangle  \nonumber \\
&=&C_{n}^{2}\left[ \alpha _{d}\left\vert \phi _{1}\left( \mathbf{r},t\right)
\right\vert ^{2}+2\times \mathrm{Re}\left( \beta _{d}\phi _{1}^{\ast }\left(
\mathbf{r},t\right) \phi _{2}\left( \mathbf{r},t\right) \right) +\gamma
_{d}\left\vert \phi _{2}\left( \mathbf{r},t\right) \right\vert ^{2}\right] ,
\label{ndensity}
\end{eqnarray}%
where the coefficients are
\begin{eqnarray}
\alpha _{d} &=&\sum\limits_{i=0}^{N_{2}}\frac{N_{2}!\left( N_{1}+i-1\right)
!N_{1}\left( 1-\left\vert \zeta \right\vert ^{2}\right) ^{N_{2}-i}\left\vert
\zeta \right\vert ^{2i}}{i!i!\left( N_{1}-1\right) !\left( N_{2}-i\right) !},
\label{ad} \\
\beta _{d} &=&\sum\limits_{i=0}^{N_{2}-1}\frac{N_{2}!\left( N_{1}+i\right)
!\left( 1-\left\vert \zeta \right\vert ^{2}\right) ^{N_{2}-i-1}\left\vert
\zeta \right\vert ^{2i}\zeta }{i!\left( i+1\right) !\left( N_{1}-1\right)
!\left( N_{2}-i-1\right) !},  \label{bd} \\
\gamma _{d} &=&\sum\limits_{i=0}^{N_{2}-1}\frac{N_{2}!\left( N_{1}+i\right)
!\left( 1-\left\vert \zeta \right\vert ^{2}\right) ^{N_{2}-i-1}\left\vert
\zeta \right\vert ^{2i}}{i!i!N_{1}!\left( N_{2}-i-1\right) !}.  \label{cd}
\end{eqnarray}%
In addition, the normalization constant is determined by%
\begin{equation}
C_{n}^{2}\left( \sum\limits_{i=0}^{N_{2}}\frac{N_{2}!\left( N_{1}+i\right)
!\left( 1-\left\vert \zeta \left( t\right) \right\vert ^{2}\right)
^{N_{2}-i}\left\vert \zeta \left( t\right) \right\vert ^{2i}}{%
i!i!N_{1}!\left( N_{2}-i\right) !}\right) =1.  \label{normconst}
\end{equation}

The above density expectation value is obtained based on the second
quantization method. Although the expressions of the coefficients given by
Eqs. (\ref{ad}), (\ref{bd}), and (\ref{cd}) are quite different from the
results calculated from the many-body wavefunction (See Ref. \cite{Xiong}),
we have proven that this density expectation value given by Eq. (\ref%
{ndensity}) is equal to the result calculated from the many-body
wavefunction $\Psi _{N_{1}N_{2}}\left( \mathbf{r}_{1},\cdots ,\mathbf{r}%
_{N_{1}+N_{2}},t\right) $ which satisfies the exchange symmetry of identical
bosons in Ref. \cite{Xiong}.

For two independent ideal condensates, before the overlapping of the two
condensates, $\zeta \left( t=0\right) =0$. Based on the Schr\H{o}dinger
equation, it is easy to verify that after the double-well potential
separating two condensates is removed, one has $\zeta \left( t\right) =0$ at
any further time. Thus $\beta _{d}=0$, and the density expectation value
given by Eq. (\ref{ndensity}) is equal to the result given by Eq. (\ref%
{nd-field(2)}). In this case, the interference term is zero in the density
expectation value.

In the presence of interatomic interaction, we now turn to investigate the
evolution equations of $\phi _{1}$ and $\phi _{2}$. The overall energy of
the whole system is
\begin{equation}
E=\int dV\left\langle N_{1},N_{2},t\right\vert \left( \frac{\hbar ^{2}}{2m}%
\nabla \widehat{\Psi }^{\dag }\cdot \nabla \widehat{\Psi }+V_{ext}\widehat{%
\Psi }^{\dag }\widehat{\Psi }+\frac{g}{2}\widehat{\Psi }^{\dag }\widehat{%
\Psi }^{\dag }\widehat{\Psi }\widehat{\Psi }\right) \left\vert
N_{1},N_{2},t\right\rangle ,  \label{overallenergy}
\end{equation}%
where $V_{ext}$ is the external potential and $g$ is the coupling constant.
By using the ordinary action principle and the above interaction energy, one
can get the following coupled evolution equations for $\phi _{1}$ and $\phi
_{2}$:
\begin{eqnarray}
i\hslash \frac{\partial \phi _{1}}{\partial t} &=&\frac{1}{N_{1}}\frac{%
\delta E}{\delta \phi _{1}^{\ast }},  \nonumber \\
i\hslash \frac{\partial \phi _{2}}{\partial t} &=&\frac{1}{N_{2}}\frac{%
\delta E}{\delta \phi _{2}^{\ast }},  \label{evo-equation}
\end{eqnarray}%
where $\delta E/\delta \phi _{1}^{\ast }$ and $\delta E/\delta \phi
_{2}^{\ast }$ are functional derivatives. With the above coupled equations,
one can understand that $\zeta \left( t\right) $ becomes nonzero after the
overlapping between two condensates for $g$ being nonzero. The reason that $%
\zeta \left( t\right) $ becomes nonzero is especially due to the nonlinear
interaction of the whole system.

Although generally speaking, $\left\vert \zeta \left( t\right) \right\vert $
is much smaller than $1$ because $\phi _{1}\phi _{2}^{\ast }$ is an
oscillation function about the space coordinate, nevertheless, a nonzero
value of $\zeta \left( t\right) $ will give significant contribution to the
density expectation value for large $N_{1}$ and $N_{2}$. It is found that
for $\left\vert \zeta \right\vert $ being the order of $N_{1}^{-1}$ and $%
N_{2}^{-1}$, the interference term in Eq. (\ref{ndensity}) will play very
important role in the density expectation value \cite{Xiong}. The coherence
effect of two initially independent condensates can be shown through the
value of $\left\vert \beta _{d}/\alpha _{d}\right\vert $. For $\beta
_{d}/\alpha _{d}=0$, there is no coherence between two condensates even
there is an overlapping. For $\left\vert \beta _{d}/\alpha _{d}\right\vert
=1 $, the two condensates can be regarded to be fully coherent.

We see that interatomic interaction plays an essential role in the emergence
of the interference effect for two initially independent condensates. In the
presence of interatomic interaction, the two initially independent
condensates will become coherent after the overlapping between two
condensates. This interference effect is a natural result of the commutation
relation $\left[ \widehat{a}_{1},\widehat{a}_{2}^{\dag }\right] =\zeta
^{\ast }$ which is nonzero for $\zeta ^{\ast }$ being nonzero in the
presence of interatomic interaction. Generally speaking, increasing the
particle number will enhance the interference effect in the density
expectation value. Based on Eq. (\ref{evo-equation}), increasing the
coupling constant $g$ will have the effect of increasing $\zeta \left(
t\right) $. Thus, increasing the interaction between particles will enhance
the coherence effect. To show more clearly the interaction-induced coherence
process between two initially independent condensates, here we will
investigate the density expectation value for different coupling constants.
In the last few years, the rapid experimental advances of Feshbach resonance
\cite{Fesh} where the scattering length can be tuned from positive to
negative make this sort of experiment be possible.

\subsection{\protect\bigskip Evolution of the density expectation value for
different coupling constants}

To give a clear presentation, we consider the evolution of the density
expectation value for one-dimensional case. At $t=0$, to give a general
comparison for different coupling constants, the initial wavefunctions for
two independent condensates are assumed as
\begin{eqnarray}
\phi _{1}\left( x_{l},t=0\right) &=&\frac{1}{\pi ^{1/4}\sqrt{\Delta _{1}}}%
\exp \left[ -\frac{\left( x_{l}-x_{1}\right) ^{2}}{2\Delta _{1}^{2}}\right] ,
\label{inital1} \\
\phi _{2}\left( x_{l},t=0\right) &=&\frac{1}{\pi ^{1/4}\sqrt{\Delta _{2}}}%
\exp \left[ -\frac{\left( x_{l}-x_{2}\right) ^{2}}{2\Delta _{2}^{2}}\right] .
\label{inital2}
\end{eqnarray}%
In the above wavefunction, we have introduced a dimensionless variable $%
x_{l}=x/l$ with $l$ being a length. In the present work, we assume that $%
\Delta _{1}=\Delta _{2}=0.5$, and $x_{2}-x_{1}=4.5$. For these parameters,
at $t=0$, the two condensates are well separated. In the numerical
calculations of the coupled equations given by Eq. (\ref{evo-equation}), it
is useful to introduce the dimensionless variable $\tau =E_{l}t/\hslash $
with $E_{l}=\hbar ^{2}/2ml^{2}$ and dimensionless coupling constant $%
g_{l}=N_{1}g/E_{l}l$.\ In addition, the particle number is assumed as $%
N_{1}=N_{2}=1.0\times 10^{5}$. In real experiments, interatomic interaction
will play very important role in the ground-state wavefunction of the
condensates confined in the double-well trap. However, in principle, one can
prepare the state given by Eqs. (\ref{inital1}) and (\ref{inital2}) by
adjusting the double-well trap for different coupling constants. In the
present theoretical work, the identical initial wavefunction for different
coupling constants will be helpful in the comparison of the density
expectation value for different coupling constants.

\begin{figure}[tbp]
\includegraphics[width=0.6\linewidth,angle=270]{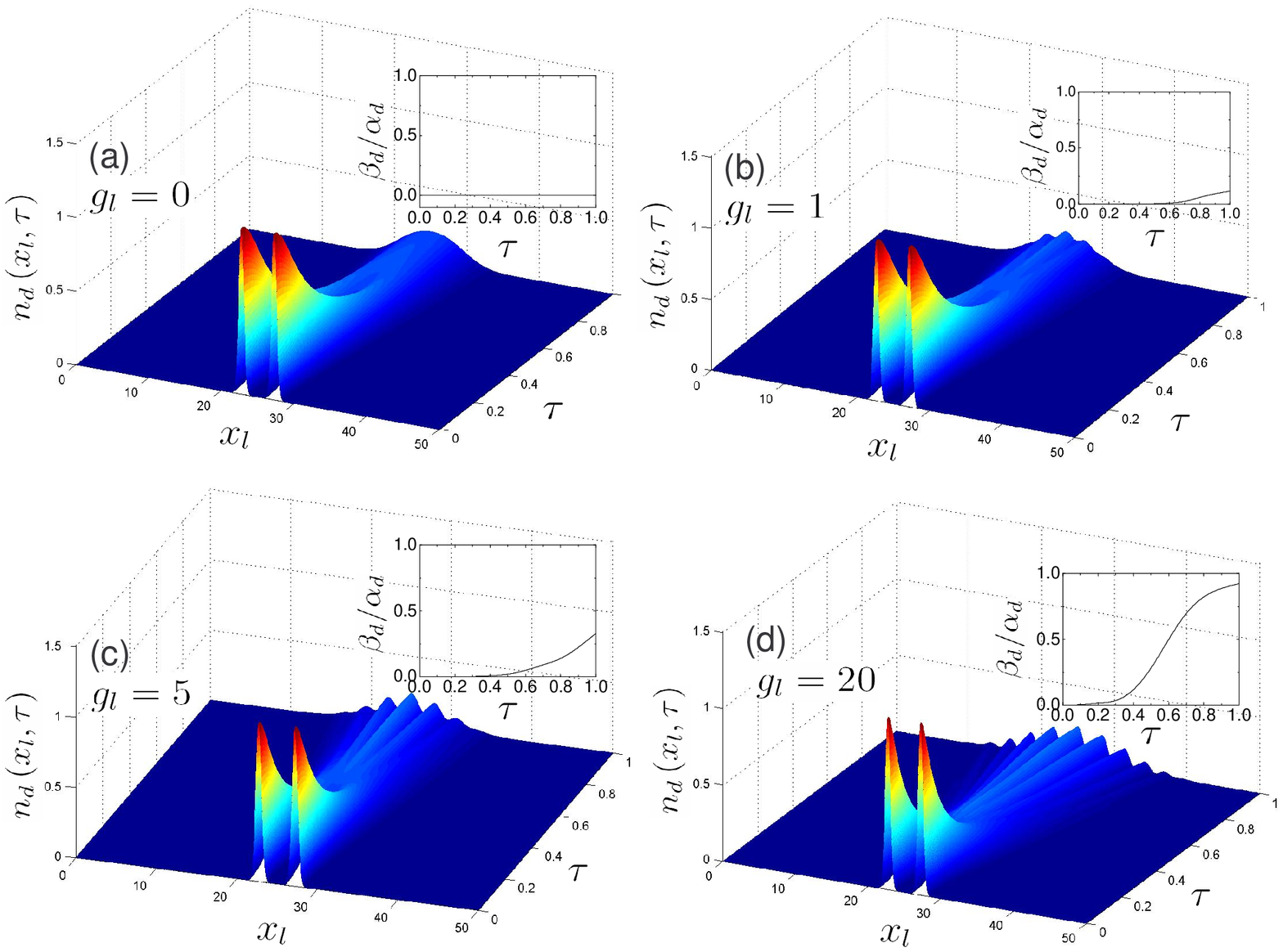}
\caption{After two initially independent condensates are allowed
to expand freely, shown is the evolution of the density
expectation value $n_{d}\left( x_{l},\tau \right) $ (in unit of
$N_{1}+N_{2}$) for different coupling constants $g_{l}$. Shown in
the inset of each figure is the relation between $\beta
_{d}/\alpha _{d}$ and dimensionless time $\tau $. For two ideal
condensates shown in figure 2(a), we see that there is no
interference pattern even there is an overlapping between two
condensates. For the case of $g_{l}=1$ shown in figure 2(b), we
see that low-contrast interference patterns begin to emerge due to
the interaction-induced coherence process. In figures 2(c) and
2(d), we see that there are high-contrast interference patterns.
In particular, in figure 2(d) for $g_{l}=20$, two initially
independent condensates can be regarded to be fully coherent near
$\tau =1$.}
\end{figure}

With these parameters, one can get the evolution of $\phi _{1}$ and $\phi
_{2}$ based on the numerical calculations of Eq. (\ref{evo-equation}) after
the double-well potential is switched off. From $\phi _{1}$ and $\phi _{2}$,
we can get $\zeta $, and thus the density expectation value based on Eq. (%
\ref{ndensity}). Shown in figure 2 is the evolution of the density
expectation value for different coupling constants. It is shown clearly that
increasing the coupling constant will enhance the coherence effect.

\subsection{Summary and discussion}

In summary, by using the second quantization method, we give the exact
expression of the density expectation value when two initially independent
condensates are allowed to overlap. The exact expression of the density
expectation value is the same as the result calculated from the many-body
wavefunction \cite{Xiong}. In the presence of interatomic interaction, the
wavefunctions of two condensates will become non-orthogonal after the
overlapping. For large particle number, the non-orthogonal property of the
wavefunctions plays very important role in the interference effect between
two condensates. Through the numerical calculations of the density
expectation value of the whole system for different coupling constants, we
show the interaction-induced coherence process between two initially
independent condensates.

\begin{acknowledgments}
This work is supported by NSFC under Grant Nos. 10474117, 10205011 and
10474119 and NBRPC under Grant Nos. 2005CB724508 and 2001CB309309.
\end{acknowledgments}

\end{document}